\documentstyle[11pt]{article}
 
\begin{document}

\title{{\bf A Monte Carlo analysis of
percolation of
 line-segments on a square lattice.}}
 \author{{\bf Y. Leroyer and E. Pommiers}\\[1cm]
{\em Laboratoire de Physique
Th\'eorique} \\ {\em CNRS, Unit\'e Associ\'ee 764}\\
{\em Universit\'e de Bordeaux I}\\
{\em 19 rue du Solarium, 33175 Gradignan Cedex, France}}
\date{}
\begin{titlepage}
\maketitle
\thispagestyle{empty}

\begin{abstract}
We study the percolative properties of bi-dimensional systems
generated by
a random sequential adsorption  of line-segments on a square lattice.
As the segment length grows, the percolation threshold decreases,
goes through a minimum and then increases slowly for large segments.
We explain this non-monotonic behaviour by a structural change
of the percolation clusters. Moreover, it is strongly suggested that
these systems do not belong to the universality class of random site
percolation. \end{abstract}
\vfill
LPTB 93-21\\
December 1993
e-mail : {\bf Leroyer@frcpn11.in2p3.fr}
\end{titlepage}

\section{Introduction}
The random site percolation model has been extensively
studied~\cite{Stauffer,Kinzel}
 and some exact results have been obtained concerning
the geometrical phase transition that this system undergoes. For most
real percolating systems, the particles are  extended objects, more or
less isotropic, and some of their physical properties  depend on the
detailed geometry of the particles. These systems form a larger class
of
models, the correlated site percolation model.  On the basis of
renormalisation-group arguments, it is believed that, as long as the
sites are correlated on a {\em finite} range, the universal properties
of these
systems are described by the random site percolation
critical point.
\cite{Stauffer,Kinzel,SE,Naka,Carmona,Boissonade}
 However, some non-universal parameters, such as
 the percolation threshold, are of physical interest, and
their determination provides useful information on the system.
Among the many authors who have investigated the various aspects of the
discrete or continuous versions of this model, let us cite the
comprehensive numerical work of Pike and Seager~\cite{PS} on the
continuous percolation problem for various extended objects.

One of these sytems, the percolation of rod-like objects, has
received much
attention due to its connection with  composite materials made of
conducting fibers embedded in an insulating
 resin~\cite{Carmona,Boissonade,Balberg}.
The percolation transition is linked to the metal-insulator transition,
and the critical concentration is an important parameter which
characterises the properties of the material. Alternatively,
the same system restricted to two dimensions, can describe
 the adsorption of conducting rod-like polymers on a
silicon substrate on which active sites are regularly
disposed~\cite{Aime}.
These systems involve very asymmetric objects, and the excluded
volume effect,
which drives the deposition process,  strongly influence the
geometry of the percolative configurations, especially in
two dimensions, and consequently, may affect the non-
universal as well as the universal
  properties of the system.

  In this paper, we investigate this system in the framework
  of a Monte-Carlo analysis of a
two-dimensional lattice model, where the rod-like objects are
simulated by line segments and the configurations are  generated by
a random sequential adsorption process with hard core exclusion.
 In the next section,
we present the technical details of the Monte-Carlo simulation,
and the determination of the
percolation threshold as a function of the segment length. The
third section is devoted to a test of the universality of this
system, and we expose our conclusions in the last section.

\section{The percolation threshold}
The model that we investigate in this paper, is defined as
follows~:
the substrate is a square periodic lattice of linear size $L$ on
which
 we drop a line segment which covers exactly $k$ sites, aligned
along one of the lattice axes, and such that the  segment sites
exactly fit the lattice. It  sticks if it does'nt overlap
previously deposited segments,
otherwise it is rejected. The process is repeated, and the density
of
the  configurations increases, until percolation occurs~:
we say
that the system percolates when there is a connected path of
occupied sites which spans the lattice either horizontally or
vertically (rule $R_0$ of ref.~\cite{RSK}).
 We denote by $p_c(L)$ the critical
concentration, averaged over
a sample of $N_S$ percolating configurations (we take $N_S$ in the
range (100, 500)).
So defined, the finite size percolation threshold $p_c(L)$
corresponds to the $p_{av}$ of
ref.~\cite{Stauffer} or $\langle p \rangle$ of ref.~\cite{RSK}, i.e.
$$p_c(L)=\int_0^1 \frac{dR_L}{dp}\; p\; dp$$
where $\frac{dR_L}{dp}dp$ is the probability for a system of size
$L$ to percolate at a density  in the range $(p, p+dp)$.\\
For each segment length $k$, we have measured $p_c(L)$ for several
(6 to 8) increasing
lattice sizes $L$ up to a value $L_{\rm max}$ such that
 $25k\le L_{\rm max}\le 1024$.
We show in table I a significant subset of our results, where
the error bars for finite $L$ are statistical.\\

We then perform the $L\rightarrow\infty$ extrapolation
by means of the  finite size scaling law~:
$$p_c=p_c(L)-AL^{-\omega}$$
If the transition belongs to the random site
percolation universality class, the exponent $\omega$ is linked to
the exponent $\nu  = 3/4$ of the correlation length~: on the basis
of the renormalisation-group, one expects
$\omega=1/\nu$~\cite{Stauffer,RSK}; however, this result has
recently been discussed  by R.M. Ziff~\cite{Ziff}, who found instead
$\omega=1+1/\nu = 7/4$.
 We have performed a fit of $p_c(L)$
according to the above equation, leaving the exponent $\omega$ as a
free parameter. The resulting $p_c(L=\infty )$ values are displayed
in
the last column of
table I and are determined with a good accuracy. In all cases,
the value of the exponent
$\omega$ is,  within the error bars, more compatible with 0.75
than with
1.75, but the precision is too poor for a firm discrimination.
\\

The extrapolated percolation threshold $p_c$, plotted in figure 1
as a function of $k$, first decreases rapidly , then flattens out and
finally begins to increase slowly for
$k\ge 10$. This effect has previously been mentionned in
ref.~\cite{Carmona}
and has been observed in a system of correlated site
percolation~\cite{SE}.
We shall justify this rather unexpected behaviour by a change of the
                            geometry
of the configurations due to excluded volume effects.

This effect is visible in figure 2 which shows the percolating
cluster at $p_c(L)$ for $k=8$ (upper figure) and for $k=32$ (lower
figure) on the same lattice of size $L=512$ ~: for large
$k$ ($k=32$), the percolation cluster exhibits dense areas of aligned
segments connected by low density regions, whereas the $k=8$ cluster
is much more uniformly distributed. In order to establish  this
effect more quantitatively, let us define the degree of alignment of
the
configuration  in the following way~:
cover a given percolating configuration with a set of square boxes of
size $\ell$; let $n_H$ be the number of sites covered by  horizontal
segments in a given box  and $n_V$  the number of sites
covered by vertical segments in the same box;  the degree of
alignment is given by $$A(\ell )=\left\langle\frac{|n_H-
n_V|}{n_H+n_V}\right\rangle$$
where the average is taken over the boxes which cover the
configuration and over a sample of different configurations.
Plotted in figure 3
as a function of  $\ell$ for
different values of $k$, this quantity slowly decreases down to 70\%
for box sizes
$\ell\simeq k$ and then falls rapidly like $1/\ell$ (dashed line) as
expected from elementary statistics for a disordered system . This
behaviour confirms the existence of
regions of same orientation and whose typical size is the segment
length $k$.\\

Assuming that the percolating cluster is an agregate of such regions,
the following argument explains qualitatively the behaviour of the
percolation threshold as a function of $k$.
Let us assimilate a typical region of oriented segments
to a  rigid square box  of size $2k$ and consider
 in it, a percolating cluster of $n$ segments ($n\le 2k$). This
cluster includes at least one segment originating at the bottom line
of the box and at least one segment ending at the top line. The $n-2$
remaining
segments ensure the connectivity, but their origins are randomly
distributed among the $k$ allowed sites, giving to this cluster a
statistical weight proportionnal to $k^{n-2}$. Its density is
$nk/(2k)^2$, and decreases like $1/k$ as long as  $n\ll 2k$, and is
constant for $n\sim k$.
Therefore, as $k$ increases,
the relative weight of the dense clusters grows rapidly, and
eventually the configurations are dominated by those clusters
whose  density remains constant with $k$. Assuming that
the variations of the percolation threshold are driven by such a
mechanism, one expects a $1/k$ decrease in the small $k$ regime,
corresponding to the suppression of the light clusters, ending
with a constant behaviour
when the configurations become dominated by denser clusters. We
have fitted our data according to this behaviour (continuous line
in figure 1), and the agreement is quite good in the small $k$
regime\footnote{A pure $1/k$ decrease has been observed in
ref.~\cite{Carmona} and explained by similar arguments}. However,
this modelisation of the oriented region by
rigid boxes is presumably too simple to explain
the slight rise  for large $k$, which may be due to the
interpenetrability of these regions and/or to their anisotropy.

This argument explains the change of geometry of the configurations
observed in figure 2 - rather diffuse for small $k$ and denser for
larger $k$ -
as well as the stopping of the $1/k$  decrease of the percolation
threshold as the segment length grows.

\section{Non-universal exponent}
The geometry of the percolative clusters is described by several
fractal exponents, one of which is the fractal dimension of the
infinite cluster $D$, which is linked to the percolation critical
exponent ratio $\beta /\nu$ $$D=d-\frac{\beta}{\nu}$$
where $d$ is the space dimensionality.
If finite range correlated site percolation is  described by the random
site percolation critical point, one expects for this fractal
dimension the exact known value $D=91/48$~\cite{Stauffer}.
However, the results of the previous section suggest that, as the
segment length $k$ grows, the fractal dimension of the critical
cluster does'nt remain constant, but increases instead. Since
$91/48$ is close to 2, in order to detect
such a tiny effect one needs a very accurate determination of  $D$.

Let $P(k,L,p)$ be the average size of the largest
cluster on a lattice of linear size $L$, for segments of length $k$
and at a concentration $p$. According to finite size scaling, one
expects~: \begin{equation}
P(k,L,p)=L^{2-\beta /\nu }\; F\left(k/L,L/\xi\right) =
L^{2-\beta /\nu }\; F\left(k/L,L(p_c-p)^\nu\right)
\label{UU1}
\end{equation}
where $\xi \simeq (p_c-p)^{-\nu}$ is the correlation length of the
connected regions. The dependence on $k/L$ has been introduced to
take into account the different length scales of the system, but
since $k\ll L$, only the limiting value $F(0,y)$ is of interest,
which we assume to be non-zero.
According to this equation, one expects that
at $p=p_c(L)=p_c-AL^{-1/\nu}$, the function
\begin{equation}
P(k,L,p_c(L))\times L^{-2+\beta /\nu}  \label{UU}
\end{equation}
depends only on the ratio $k/L$ and tends to a non zero constant
value as $k/L \rightarrow 0$.

Alternatively, if there is not universality, the critical exponents
become $k$-dependent and one expects
\begin{equation}
P(k,L,p_c(L)) =  L^{2-\omega (k)}\; G(k/L)  \label{NU}
\end{equation}
where $\omega$ progressively deviates from
$\beta /\nu = 5/48$ as $k$ increases.

We have measured $P(k,L,p_c(L))$ for several values of $k$ as a
function of $L$ in order to discriminate between these two
different
 behaviour. This measure requires
a very accurate determination of $p_c(L)$, because $P$ varies quite
rapidly in this $p$ region, from the non-percolating regime where
the exponent
is zero, to the percolating one, where the exponent is 2. We
selected three values, $k=2,8,16$, and determined $p_c(L)$ with a
precision less than 0.2\% .
Then we proceeded to the measure of $P(k,L;p_c(L))$. In order to
take into account the systematic error induced by a small variation
of $p$,
we have performed two measures, at $p_c(L)\pm \Delta p_c(L)$, and
 the error bars displayed in the following figures represent the
                           difference
between the two results (the statistical errors are smaller than
this systematic one).

In figure 4, we have plotted the quantity $P(k,L,p_c(L))\times L^{-
91/48}$ as a function of $k/L$, for the three $k$ values.
According to eq.~\ref{UU}
one should expect all the points to lie on the same universal curve. In
spite of the rather large error bars, we see that this is not the case.
In figure 5 we display $P(k,L,p_c(L)) /L^2$ as a function of $L$
in a Log-Log plot. The straight lines correspond to the best fit
according to the power law of eq.~\ref{NU} and with the
exponents~:
$\omega (k=2)=0.120\pm 0.009$, $\omega (k=8)=0.084\pm 0.011$ and
$\omega (k=16) = 0.011^{+0.020}_{-0.011}$.
As $k$ increases, one clearly observes a variation of the exponent
$\omega (k)$, away from the universal value $\omega (k=1)=0.104$.

All these results are more compatible with
eq.~\ref{NU} than eq.~\ref{UU} and strongly suggest
that these systems are not in the random site percolation
universality class.

\section{Conclusion}
In conclusion, we have exhibited a non-monotonic behaviour of the
critical concentration for percolating systems of line segments
deposited on a square
lattice, as the segment length increases. We give a simple
argument, based on the local alignment effect, which explains the
change of structure
of the critical clusters and the non-decreasing behaviour of the
percolation threshold for large $k$.

 The observed slow increase  may have its origin in
the  effect proposed by Sanders and Evans~\cite{SE} to explain
similar behaviour in a correlated site percolation model based on an
island forming process. In the limit of infinitely attractive force,
this process favours the formation of large isotropic islands and
evolves towards the continuous percolation model of Pike and
Seager~\cite{PS},
with a threshold of 0.68, higher
than for finite attractive force. Therefore, as the force strength
increases the percolation threshold is pulled up toward this
asymptotic value.
It is tempting to establish a parallel between this model and our
system in the following (speculative) way~:   the oriented regions
of growing size play the role of the islands of the above-mentionned
process, and eventually simulate
a continuous percolation model for extended anisotropic objects.
Assuming the same critical concentration $p_c\simeq 0.68$ as for
the isotropic objects of Pike and Seager, we expect $p_c$ to increase
toward this asymptotic value as $k$ increases. It is worth
noticing that in this limit, the {\em saturation coverage} of the
system
is 0.66~\cite{YL} and the percolation concentration may well be
unreachable.\\

Another consequence of the geometrical change of the configurations
when $k$ varies, is that the various fractal exponents of the
infinite cluster may depend on $k$. Our results strongly suggest that
the fractal dimension
of the infinite cluster, linked to the ratio of the critical
exponents $\beta$ and $\nu$, slowly increases, thus
indicating for these systems a departure from the expected
random site percolation universal behaviour.

 However, this effect appears as a small signal, and in spite of
a careful analysis one cannot rule out the possibility of
systematic errors due, for instance, to some
underestimated finite size effect linked to a too small $L/k$ ratio.
Therefore these results need to be confirmed by a determination of
other
critical exponents linked to the cluster structure, such as
$\gamma$, which is characteristic of the average cluster size.
Furthermore, if a
violation of universality is definitely established, one must
explain its origin. Presumably, the anisotropy of the percolating
objects plays a role, and, for large segments, the transition may
induce a large scale anisotropy.
 In this case, one should characterise the scaling
properties of the percolating clusters by critical exponents in
both lattice directions, just like in  the directed
percolation model~\cite{Kinzel2}.
These various aspects of the model are currently under
investigation. \\

{\bf Acknowledgements}\\
We thank F. Carmona for several helpful discussions and J.T.
Donohue
for a careful  critical reading of the manuscript.
YL thanks Jim Evans for pointing out  ref.~\cite{SE} to him.

\newpage
{\footnotesize
\begin{center}
\begin{tabular}{lllllllll}
\hline \hline
&\multicolumn{7}{c}{ $p_c(L)$} & $p_c$\\
\hline
$k\setminus L$
&  128    &  256    &   384   &   512   &   680  &  850   &  1024
                          &$\infty$ \\
\hline
1 &0.5860(5)&0.5890(6)&0.5907(6)&0.5909(4)&        &        &        &
0.593(1)\\
2 &0.5551(4)&0.5577(4)&0.5587(3)&0.5596(2)&        &        &        &
0.561(1)\\
4 &0.498(2) &0.5000(9)&         &         &        &        &        &
0.504(3)\\
8 &0.4562(8) &0.4628(8)&0.4650(6)&0.4653(4)&       &        &0.4674(5)&
0.470(1)\\
16 &0.442(1) &0.4502(7) &0.4560(8) &0.4578(7)&0.459(1)&0.460(1)&0.461(1)&
0.463(1)\\ 24 &         &0.445(3) &0.458(2) &0.459(2)
&0.462(1)&0.462(1)&0.464(1)& 0.466(1)\\ 32 &0.402(3) &0.450(2) &0.462(3)
&0.466(1) &0.467(2)&0.467(1)&0.469(1)& 0.471(1)\\ 40 &         &0.450(2)
&0.461(1) &0.467(1) &0.470(2)&0.474(2)&0.475(2)& 0.484(6)\\ \hline \hline
\end{tabular}\vspace{1cm}
 \end{center}}
{\bf Table I} : The percolation threshold for various segment lengths $k$
 and lattice sizes $L$. The last column corresponds to the extrapolated
value. Numbers in parentheses are the errors affecting
 the corresponding final digits.\\

\newpage
{\Large {\bf Figure captions}}\\
\begin{itemize}
\item[]{\bf Figure 1} The extrapolated percolation threshold as a
function of the segment length $k$. The continuous line is a fit in
$1/k+$ const.
\item[]{\bf Figure 2} Two percolating clusters, for segments of length
$k=8$ and $k=32$, on a lattice of size $L=512$.
\item[]{\bf Figure 3} The degree of alignment (as defined in the text)
as a function of the block size $\ell$, for different segment lengths.
The dashed lines are fits in $1/\ell$ of the tails of the distributions.
\item[]{\bf Figure 4} The quantity $P(k,L,p_c(L))\times L^{-
91/48}$ as a function of $L/k$ for three values of the segment
length $k$. \item[]{\bf Figure 5} The quantity $P(k,L,p_c(L))/
L^2$ as
a function of $L$ for three values of the segment length $k$.
The dashed straight lines are the  power law fits.

\end{itemize}

\newpage
\begin{thebibliography}{99}
\bibitem{Stauffer}{\em Introduction to percolation theory}, D.
Stauffer, (Taylor and Francis, London,  1985)
\bibitem{Kinzel} {\em Percolation structures
and processes}, Ann. Isr. Phys. Soc. Vol. 5, edited by G. Deutsher,
R. Zallen and J. Adler, 1983.

\bibitem{SE}Sanders D.E., Evans J.W. {\em Phys. Rev.} {\bf A38},
4186 (1988) \bibitem{Naka} M. Nakamura, {\em Phys. Rev.} {\bf A36},
2384 (1987) \bibitem{Carmona}Carmona F., Barreau F., Delhaes P. and
Canet R.
{\em J. Physique Lett.} {41}, L534 (1980)
\bibitem{Boissonade} Boissonade J., Barreau F.,
Carmona F. {\em J. Phys.} {\bf A16}, 2777 (1983)
\bibitem{PS}G.E. Pike and C.H. Seager, {\em Phys. Rev.} {\bf B10},
1421 (1974) \bibitem{Balberg} I. Balberg, N. Binenbaum {\em Phys.
Rev.} {\bf B28},
3799 (1983)
\bibitem{Aime}Experiment in progress at LCPC, Bordeaux University,
J.P. Aime private communication.

\bibitem{RSK} P.J. Reynolds, H.E. Stanley and W. Klein, {\em J. Phys.}
{\bf A11},L199 (1978)\\
P.J. Reynolds, H.E. Stanley and W. Klein, {\em Phys. Rev.} {\bf B21},
1223 (1980)
\bibitem{Ziff}R.M. Ziff, {\em Phys. Rev. Lett.} {\bf 69}, 2670 (1992)

\bibitem{YL}B. Bonnier, M. Hontebeyrie, Y. Leroyer, C. Meyers and E.
Pommiers, to appear in {\em Phys.~Rev.~E}
\bibitem{Kinzel2} for a review, see the article of W. Kinzel in
ref.~\cite{Kinzel}
\end{thebibliography}
\end{document}